\documentclass[twocolumn,showpacs,preprintnumbers,amsmath,amssymb]{revtex4}

\usepackage{graphicx}
\usepackage{dcolumn}
\usepackage{bm}
\usepackage{SIunits} 

\begin{document}

\preprint{For Applied Physics Letters}

\title{Role of microstructure in porous silicon gas sensors for NO$_2$}

\author{Zeno Gaburro}
\email{gaburro@science.unitn.it}
\homepage{http://www.science.unitn.it/~semicon}
\author{Paolo Bettotti}
\author{Massimo Saiani}
\author{Lorenzo Pavesi}

\affiliation{INFM and Department of Physics,
             University of Trento, Italy}

\author{Lucio Pancheri}
\affiliation{Department of Information and Communication
             Technology, University of Trento, Italy}

\author{Claudio J. Oton}
\author{Nestor Capuj}

\affiliation{Departamento de F\'isica B\'asica,
             Universidad de La Laguna, Tenerife, Spain}

\date{\today}

\begin{abstract}
Electrical conductivity of porous silicon fabricated form heavily
doped p-type silicon is very sensitive to NO$_2$, even at
concentrations below 100~ppb. However, sensitivity strongly
depends on the porous microstructure. The structural difference
between sensitive and insensitive samples is independently
confirmed by microscopy images and by light scattering behavior. A
way to change the structure is by modifying the composition of the
electrochemical solution. We have found that best results are
achieved using ethanoic solutions with HF concentration levels
between 13\% and 15\%.
\end{abstract}



\maketitle

Porous silicon (PSi) is an interesting material for gas
sensing~\cite{1997Sailor}. Physical properties of PSi, such as DC
electrical conductivity, are very sensitive to the
environment~\cite{1997Bilenko}. Detection of NO$_2$ at
concentration levels as low as 12~ppb has been
demonstrated~\cite{2003Pancheri}. Thus, PSi is an intriguing
material for NO$_2$ sensors. NO$_2$ is a well known air pollutant,
originated by internal combustion engines, whose attention level
is set at around 100~ppb by pollution normatives.

Exposure of PSi to NO$_2$ leads to an increase of DC conductivity.
The effect has been observed on PSi fabricated from heavily doped
p-type (p$^+$) substrates (resistivity $\rho$ in the
\milli\ohm~\centi\meter{} range). High sensitivity is reported in
thick samples (at least some tens of \micro\meter), and in the
porosity range between 50\% and
80\%~\cite{2000Boarino,2001Baratto,2001Timoshenko,2001Boarino,2003Pancheri}.

Porous layers obtained from heavily p-type doped wafers have low
conductivity, even though the boron dopants are not significantly
removed during the anodization process; boron concentration is
comparable in PSi and bulk Si~\cite{1998Polisski,1999Polisski}. It
is argued that one of the relevant effects of anodization to the
conductivity is to leave the dopant ions at close distance from
the surface, where defects trap free carriers, thus inhibiting
dopants' acceptor function and lowering the porous layer
conductivity~\cite{1998Polisski,1999Polisski,2001Timoshenko,2001Boarino}.
Mobility is also lower in PSi~\cite{2001Timoshenko}. Under
exposure to NO$_2$, the hole concentration increases, thus
suggesting that the acceptor function of boron dopants is
re-activated by NO$_2$~\cite{2001Timoshenko,2001Boarino}. A
characterization of the effect at different porosity levels has
been performed~\cite{2000Boarino}. However, porosity is not an
exhaustive parameter to link the sensitivity and the
microstructure of PSi. To our knowledge, it has not been hitherto
shown whether the microstructure of PSi plays a role in the
sensitivity to NO$_2$. The aim of this work is to demonstrate that
the microstructure has a critical role in such sensitivity.

PSi layers were grown by electrochemical dissolution in an
HF-based solution on a single-crystalline p-type (100)
heavily-doped Si substrate. Substrate nominal resistivity $\rho$
was 6-15~\milli\ohm~\centi\meter. Before the anodization, the
native oxide was removed from the backside of the wafers, and
aluminium back contacts were deposited by evaporation. The
anodizing solution was obtained by mixing aqueous HF (48\% wt.)
with ethanol. We have tested different solutions, introducing
small variations in the final nominal concentration of HF, which
was ranging between 13\% and 15\% wt. As expected, lower (higher)
HF concentration led to higher (lower) porosity
samples~\cite{2000Bisi}. The etching was performed by applying an
etching current density of
50~\milli\ampere\per\centi\meter\squared{} for 23 minutes. After
anodization, the samples where rinsed in ethanol and pentane, and
dried in ambient air. Gold electrodes were deposited by
evaporation on the PSi top surface. Special care was taken to
achieve the same electrode size on all the samples. Copper wires
were connected to the gold electrodes using an epoxy silver paste.
Thickness and refractive index of samples were extracted from
normal reflectance spectra and Scanning Electron Microscopy (SEM)
images. Using Bruggeman approximation we have estimated the
porosity from the measured refractive index~\cite{2000Bisi}.
Microstructure was characterized with Transmission Electron
Microscopy (TEM) measurements.

During measurements in presence of NO$_2$ and water vapor, the
sensors were biased between one of the top contacts and the back
contact at a constant voltage, while the current was measured. The
sensors were kept in a sealed chamber under controlled flux of
gases coming from certified cylinders. Humid air was obtained by
flowing dry air through a bubbler. Different relative humidity
levels and NO$_2$ concentrations were obtained mixing humid air,
dry air and a dilute solution of NO$_2$ in air (550~ppb) with a
flow control system. Relative humidity was monitored using a
calibrated hygrometer.

\begin{figure}
  \includegraphics[scale=0.5,clip]{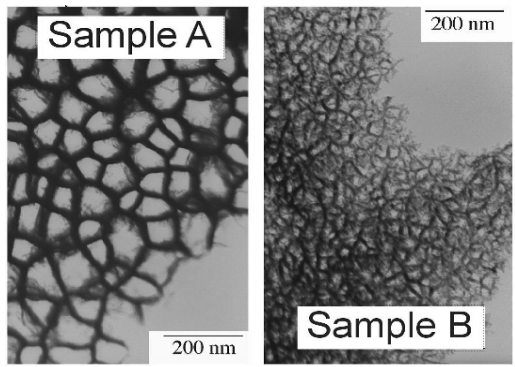}\\
  \caption{TEM images of top view (100 plane) of two porous
           silicon samples. Sample~A: porosity (extracted from
           reflectance)~=~78\%,
           thickness~=~32.5~\micro\meter.
           Sample~B: porosity (extracted from reflectance)~=~60\%,
           thickness~=~37.2~\micro\meter.
           }\label{TEM}
\end{figure}
Figure~\ref{TEM} shows the TEM images of two samples. The porosity
of Sample A and B was, respectively, 78\% and 60\%. In Figure 2,
we show the effect of exposure to water vapor (at 30\% and 70\%
levels of relative humidity) and to NO$_2$ at 50~ppb
concentration.
\begin{figure}
  \includegraphics[scale=1,clip]{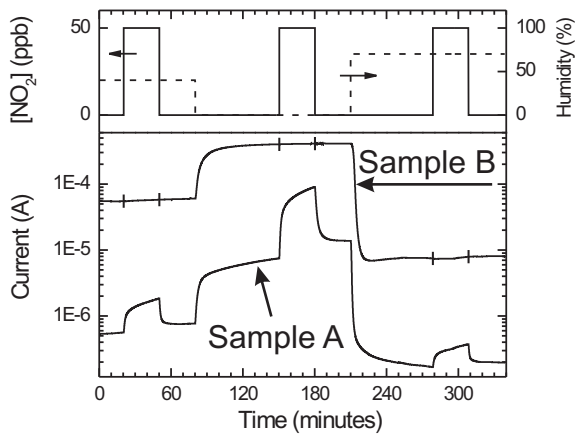}\\
  \caption{Simultaneous measurement of electrical conduction of
 Samples~A and~B under controlled atmosphere.
 The top graph show
 the composition of the gas. Solid line: NO$_2$ concentration,
 either 0 or 50~ppb, left axis. Dashed line:
 relative humidity (40\%, 0 or 70\%, right axis).
 Bottom graph: electrical current under DC
 constant voltage bias (1 V) between the
 top gold electrode and the p$^+$ substrate, during
 exposure to gas. NO$_2$ has no distinguishable effect
 on Sample B (transversal markers across Sample B
 data indicate the NO$_2$ on and off switching points).
}\label{sensing}
\end{figure}

In Figure~\ref{sensing}, no resistivity change is observed in the
60\% porosity sensor (Sample B) under exposure to NO$_2$ (50~ppb),
as opposed to the 78\% porosity sensor (Sample A). One might be
initially tempted to conclude that higher porosity leads to higher
sensitivity to NO$_2$. However, a different conclusion would be
drawn from other reports, in which samples with 60\% porosity
showed higher sensitivity than samples with 75\% porosity (the
thickness and substrate resistivity were comparable to this
work)~\cite{2000Boarino}.

The peculiarity of our fabrication procedure is low HF
concentration. At the HF concentration and the current density of
this work (respectively, 13\%-15\% and
50~\milli\ampere\per\centi\meter\squared), the anodization is
close to the electropolishing regime~\cite{2000Bisi}. In these
conditions, and especially at the lower HF concentration,
anodization is quite aggressive. In p$^+$ samples, the etching
should evolve selectively \emph{around} the dopants, leaving them
in place~\cite{1999Polisski}. However, during anodization of
Sample~A, several boron dopants have been removed. In fact, the
boron density of these samples
(N$_A\simeq10^{19}$\per\centi\meter\cubed) implies the presence of
about 1 ion every 2 or 3~\nano\meter{} along any linear direction,
whereas Sample~A has empty gaps of several tens of \nano\meter{}
(Figure~\ref{TEM}).

Changes in the solutions in aggressive conditions lead to
differences in conductivity which go beyond the difference of
porosity. For example, from the porosity difference, the amount of
leftover Si in Sample A (78\% porosity) is about two times less
than in Sample B (60\% porosity). One might expect the resistance
of the two samples to differ by a comparable factor. However, the
resistance is 2 orders of magnitude larger in Sample A
(Figure~\ref{sensing}). A possible relevance of quantum
confinement effects can be probably excluded, considering that the
cross sections of the resistive paths appear to have diameter
larger than a few \nano\meter{} (Figure~\ref{TEM}).
Figure~\ref{TEM} might even suggest that Sample B has conductive
channels with smaller cross sections. In this latter case, any
relevance of the size of the cross section would make the 2 orders
of magnitude difference in conductivity even more surprising,
because it could only increase the effective resistivity of Sample
B.

\begin{figure}
  \includegraphics[scale=0.3,clip]{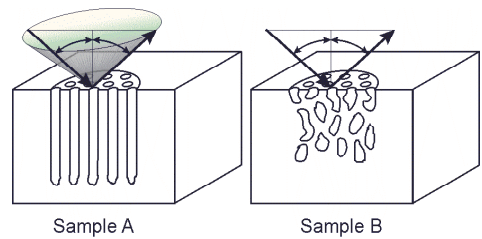}\\
  \caption{Light reflection and scattering by Samples~A and~B.
           Incident (reflected) light beams are
           represented by bold, downward (upward)
           arrows. In Sample~A, strong light
           scattering is also observed.
           Scattered rays lay on the cone generated
           by the direction of
           incidence. No such scattering is observed
           in Sample~B.
           }\label{gonna}
\end{figure}

Beside porosity, a structural difference is apparent by comparing
the TEM images. The Si structures of Sample B are more branching
and interconnecting with each other than those of Sample A. This
is confirmed both by side-views TEM images of pore walls (not
shown), and by light scattering experiments (Figure~\ref{gonna},
experimental data shown elsewhere~\cite{2002Oton,2003Oton}).
The light scattering behavior cannot be discussed on the only
basis of porosity difference. On the contrary, the scattering of
Sample A can be quantitatively explained as a structural feature.
It is originated by the straight pore walls, as discussed in
greater detail elsewhere~\cite{2002Oton,2003Oton}. The markedly
different light scattering behavior strongly emphasizes the
microstructural difference between the two samples.

Aggressive anodization leads to high sensitivity to NO$_2$
(Figure~\ref{sensing}). We propose the following connection
between microstructure and sensitivity.
\begin{figure}
  \includegraphics[scale=0.3,clip]{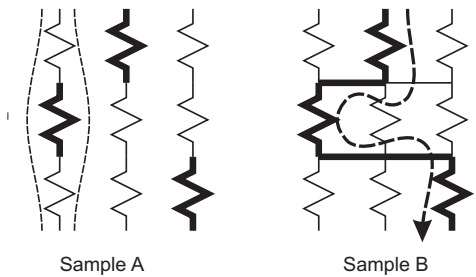}\\
  \caption{Schematic interpretation of resistive paths of samples
           shown in Figure~\ref{TEM}. Larger (smaller) resistance
           is represented by thin (thick) resistors.  In Sample A,
           paths are less interconnected, and high resistors represent
           the conduction bottleneck. Local increases of resistance
           can be due, for example, to wall narrowing (undulating
           shape, thin dashed lines at left side of Sample~A)
           and/or to higher dopant inhibition. In the more
           interconnected mesh (Sample B), high resistors are more
           likely to be bypassed.
           The selection of a
           dominating low resistance path is emphasized by a dashed
           line with arrow.
           }\label{schematic}
\end{figure}
The resistors of Figure~\ref{schematic} represent local resistance
along pore walls. Pore walls of Sample A are less interconnected,
thus high resistance portions are less likely to have local
bypasses with lower resistance. Therefore, high resistivity paths
dominate in the total Sample A resistance. The highly resistive
pore walls are the most sensitive to NO$_2$, since thinner walls
have larger fraction of dopants at close distance from the
surface. NO$_2$ locally reactivates the acceptor dopant, thus
reverting the high resistivity to lower resistivity paths. In the
case of Sample B, the presence of bypasses, whose resistance is
low even in absence of NO$_2$, obscures the effect of NO$_2$. This
interpretation qualitatively agrees with both the large difference
of resistance and the difference of sensitivity to NO$_2$.

The fact that both sensors are similarly sensitive to water
probably depends on the much larger amount of water molecules,
which acts as donors~\cite{1987Kiselev} and tend to increase the
resistance in all the porous layer.

In conclusion, DC electrical conductivity of heavily doped porous
silicon can be very sensitive to NO$_2$. The comparison of our
results with past literature shows that there is no univocal
relationship between sensitivity and porosity. In this work, we
have compared light scattering, TEM images, and sensing
performance of porous silicon samples. Our results suggest that
the structure of porous silicon is determinant towards high
sensitivity to NO$_2$. The microstructure depends on the
composition of the electrochemical solution used for the
anodization. A way to achieve very sensitive structures to NO$_2$
is by using electrochemical solutions with low HF concentrations
($\simeq$13\%). The detectable level of NO$_2$ in air is well
below 100~ppb, the threshold for realistic applications.

We acknowledge the help of Prof. S. Gialanella (Facolt\`a di
Ingegneria, Universit\`a di Trento) for the TEM measurements, the
support of INFM, progetto PAIS 2001 "SMOG", of Provincia Autonoma
di Trento and of Science and Technology Ministry of Spain (MCYT)
(Project No. MAT 2002-00044). C. O. acknowledges University of La
Laguna and Cajacanarias for the fellowship "Beca de
investigaci\'on para doctorandos" (2003).

\bibliography{2003_structure}

\end{document}